\documentclass[aip,
amsmath,amssymb,
preprint,%
numerical,
]{revtex4-1}
\usepackage{graphicx}% Include figure files
\usepackage{dcolumn}% Align table columns on decimal point
\usepackage{bm}% bold math
\usepackage{hyperref}
\usepackage[utf8]{inputenc}
\usepackage[T1]{fontenc}
\usepackage{mathptmx}
\usepackage{etoolbox}
\newcommand{\RNum}[1]{\uppercase\expandafter{\romannumeral #1\relax}}
\usepackage{array,xcolor,colortbl}
\usepackage{multirow}
\makeatletter
\def\@email#1#2{%
	\endgroup
	\patchcmd{\titleblock@produce}
	{\frontmatter@RRAPformat}
	{\frontmatter@RRAPformat{\produce@RRAP{*#1\href{mailto:#2}{#2}}}\frontmatter@RRAPformat}
	{}{}
}%
\makeatother
\begin{document}

	\title{Rotation and Oblique Irradiation Effects on Phototactic Algal Suspension Instability}
	% Force line breaks with \\

		% Force line breaks with \\
	\author{S. K. Rajput}
	\altaffiliation[Email: ]{shubh.iitj@gmail.com}
	%\author{M. K. Panda}%
	%\email{mkpanda@iiitdmj.ac.in}
	%\email{shubh.iiitj@gmail.com}
	\affiliation{ 
		Department of Mathematics, PDPM Indian Institute of Information Technology Design and Manufacturing,
		Jabalpur 482005, India.%\\This line break forced with \textbackslash\textbackslash
	}%
	%\homepage{http://www.Second.institution.edu/~Charlie.Author.}
	%\affiliation{%
		%	Second institution and/or address%\\This line break forced% with \\
		%}%
	

\begin{abstract}
	In this study, we aim to explore the behavior of microorganisms in response to natural lighting conditions, considering the off-normal angles at which the sun strikes the Earth's surface. To achieve this, we investigate the effect of oblique irradiation on a rotating medium, as this combination represents a more realistic scenario in the natural environment. Our primary focus is on understanding the phototactic behavior of microorganisms, which refers to their movement towards or away from light. Under conditions of low light, microorganisms tend to exhibit positive phototaxis, moving towards the light source, while in intense light, they display negative phototaxis, moving away from the light source. By studying a suspension that is illuminated by oblique collimated flux with a constant radiation intensity applied to the top surface, we can gain insights into how microorganisms respond to varying light conditions and rotation. The stability analysis is conducted using linear perturbation theory, which allows us to predict both the stationary and oscillatory characteristics of the bio-convective instability at the onset of bioconvection. Through this analysis, we observe that rotation plays a significant stabilizing role in the system, while oblique irradiation has a destabilizing effect on the suspension.

\end{abstract}

\maketitle
\section{INTRODUCTION}

     Bioconvection, a captivating phenomenon, showcases the convective motion observed in fluid containing self-propelled motile microorganisms like algae and bacteria at a macroscopic level ~\cite{20platt1961,21pedley1992,22hill2005,23bees2020,24javadi2020}. These microorganisms exhibit an intriguing tendency to move upwards on average due to their higher density compared to the surrounding medium, typically water. The formation of distinct patterns in bioconvection is closely tied to the behavior of these motile microorganisms. However, pattern formation is not solely reliant on their upswimming or higher density; rather, it is influenced by their response to various environmental stimuli known as "taxes," including gravitaxis, chemotaxis, phototaxis, and gyrotaxis. In this article, our focus centers on exploring the impacts of phototaxis.

     Experimental studies have provided significant insights into the influence of oblique collimated flux on bioconvective patterns~\cite{1wager1911,2kitsunezaki2007}. The intensity of light plays a crucial role in shaping stable patterns observed in suspensions of motile microorganisms in well-stirred cultures. Bright light can either disrupt existing patterns or prevent pattern formation altogether, and it significantly affects the size, shape, structure, and symmetry of the patterns~\cite{3kessler1985,4williams2011,5kessler1989}.

     In this study, we delve into the captivating world of bioconvection patterns, examining the intricate interplay between light intensity and the phototactic behavior of microorganisms. To explore this fascinating phenomenon, we employ the phototaxis model proposed by S. Kumar~\cite{40kumar2023} (referred to as SK), incorporating the Navier-Stokes equations and a microorganism conservation equation. Our investigation takes place in a suspension of phototactic microorganisms, assumed to rotate around the z-axis at a constant angular velocity, while being illuminated from above with collimated irradiation. However, we recognize a critical limitation in previous studies, such as Kumar's~\cite{40kumar2023}, as they solely considered the impact of vertical collimated flux on phototaxis. In the natural world, sunlight reaches the Earth's surface at various off-normal angles, resulting in a more complex and diverse lighting scenario. Acknowledging the importance of representing real-world conditions, we expand our analysis to encompass the influence of oblique collimated flux. By incorporating the effects of this more realistic illumination, our study aims to provide a comprehensive perspective on how microorganisms respond to varying light intensities. This extension allows us to gain deeper insights into the behavior of algae cells in their natural light environment, shedding light on the intricate dynamics behind bioconvection pattern formation. Through this approach, we seek to contribute valuable knowledge to the field, bridging the gap between theoretical models and the complexities of real-world scenarios. By unveiling the impact of oblique collimated flux on phototaxis, we aim to advance our understanding of the fascinating interactions between microorganisms and light, further illuminating the captivating mechanisms that give rise to bioconvection patterns.

     Focusing on a suspension of finite depth illuminated from above, the article analyzes the balance between phototaxis due to absorption of light and diffusion caused by the random swimming motions of the cells. This balance leads to the formation of a concentrated layer of microorganisms, known as the sublayer, which is horizontally oriented. The position of the sublayer within the suspension depends on the critical light intensity ($G_c$). The article investigates the scenarios where the sublayer forms either at the mid, three-quarter, or top of the suspension by adjusting angle of incidence.

      Phototactic bio-convection (PBC) has captivated the attention of researchers, leading to extensive investigations in the scientific literature. Previous studies by Vincent and Hill~\cite{12vincent1996}, Ghorai and Hill~\cite{10ghorai2005}, and Ghorai et al.\cite{7ghorai2010} have delved into various aspects of PBC, shedding light on the impact of factors such as negative buoyancy of cells and light scattering on the suspension. Additionally, studies by Panda and Singh\cite{11panda2016}, Panda et al.\cite{15panda2016}, and Panda\cite{8panda2020} have explored the effects of diffuse and collimated irradiation on PBC, while others investigated the influence of oblique collimated irradiation~\cite{16panda2022,17kumar2022} and rotation~\cite{39kumar2023,40kumar2023,41rajput2023}. Despite the wealth of research, there remains an important knowledge gap in understanding the collective impact of rotation and oblique collimated flux on an algal suspension. In this study, we aim to bridge this gap and investigate the combined influence of rotation and oblique collimated flux on the onset of instability in the suspension.

      Divided into three primary sections, the article presents a mathematical model to study phototaxis and calculate the equilibrium state. It further analyzes the linear stability of the equilibrium state using perturbation theory and solves it numerically using the Newton-Raphson-Kantorovich method to obtain neutral curves. Finally, the authors discuss and interpret the results based on these neutral curves, shading light on the intricate behavior of phototactic microorganisms in the captivating world of bioconvection. Through this investigation, we aim to deepen our understanding of the interactions between light, microorganisms, and fluid dynamics, offering valuable insights into the intricate mechanisms governing this fascinating natural phenomenon.
	
\section{MATHEMATICAL FORMULATION}
	
 	 In this study, the focus is on the movement of a dilute phototactic algal suspension within a confined space between two parallel horizontal boundaries. The suspension is subjected to illumination from above, with collimated flux. The depth of the suspension is fixed at $H$, and the boundaries are impermeable. The light intensity at any specific location $\boldsymbol{x}$ within the suspension, in a particular unit direction $\boldsymbol{r}$, is denoted by $L(\boldsymbol{x}, \boldsymbol{r})$.
	
\subsection{The swimming orientation}
	
     The radiative transfer equation (RTE) provides a mathematical framework for describing the behavior of radiation in a medium, including an algal suspension, with absorption and scattering. It allows us to calculate the light intensity at a particular location $\boldsymbol{x}$ and in a specific direction $\boldsymbol{r}$ within the suspension. The RTE equation, in its general form, can be written as:
\begin{equation}\label{1}
	 \boldsymbol{ r}\cdot\nabla L(\boldsymbol{x},\boldsymbol{r})+a L(\boldsymbol{x},\boldsymbol{r})=0,
\end{equation}
     The absorption coefficient, denoted as, $a$ is the parameter used to describe the interaction of light with the algal suspension.

\begin{figure}[!h]
	\centering
	\includegraphics[width=14cm]{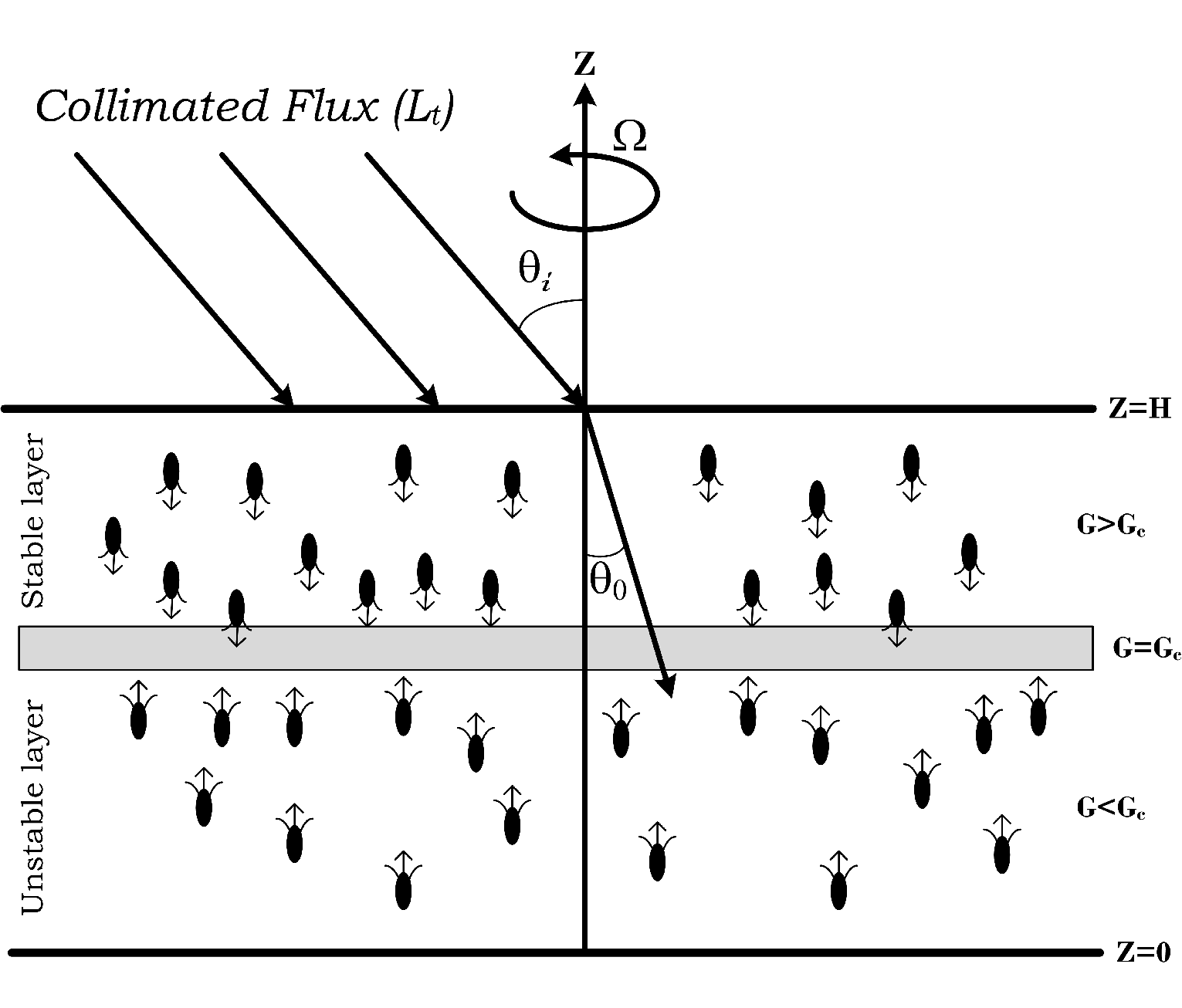}
	\caption{\footnotesize{Axial representation of the problem.}}
	\label{fig1}
\end{figure}
    The light intensity at the suspension's top can be represented as
\begin{equation*}
	L(\boldsymbol{x}_b,\boldsymbol{r})=L_t\delta(\boldsymbol{r}-\boldsymbol{r_0}) 
\end{equation*}
    where $\boldsymbol{x}_b=(x,y,H)$ is the location on the top boundary surface. Here, $L_t$ is the magnitude of oblique collimated flux. Consider $a=\alpha n(\boldsymbol{x})$. With these substitutions, Eq.~(\ref{1}) can be rewritten as
\begin{equation}\label{2}
	\boldsymbol{ r}\cdot\nabla L(\boldsymbol{x},\boldsymbol{r})+\alpha nL(\boldsymbol{x},\boldsymbol{r})=0.
\end{equation}
    The value of the total intensity at a given point $\boldsymbol{x}$ is determined by 
\begin{equation*}
	G(\boldsymbol{x})=\int_0^{4\pi}L(\boldsymbol{x},\boldsymbol{r})d\Omega,
\end{equation*}
    and similarly, the radiative heat flux is given by 
\begin{equation}\label{3}
	\boldsymbol{q}(\boldsymbol{x})=\int_0^{4\pi}L(\boldsymbol{x},\boldsymbol{r})\boldsymbol{r}d\Omega.
\end{equation}
    We assume that the cells and fluid flow at the same speed. Therefore, we can define the cells mean swimming velocity as
\begin{equation*}
	\boldsymbol{W}_c=W_c<\boldsymbol{p}>,
\end{equation*}
    here, $W_c$ denotes the cells mean swimming speed average swimming speed, and $<\boldsymbol{p}>$ denotes the mean direction of the cell's swimming which is determined by using the following equation
\begin{equation}\label{4}
	<\boldsymbol{p}>=-M(G)\frac{\boldsymbol{q}}{|\boldsymbol{q}|}.
\end{equation}
    Here, taxis response function (taxis function) $M(G)$ describes how algae cells react to light and take a mathematical form as 
\begin{equation*}
	 M(G)=\left\{\begin{array}{ll}\geq 0, & \mbox{ } G(\boldsymbol{x})\leq G_{c}, \\
		< 0, & \mbox{ }G(\boldsymbol{x})>G_{c}.  \end{array}\right. 
\end{equation*}
     When the light intensity reaches a critical value ($G = G_c$), the microorganisms exhibit zero mean swimming direction. The form of the taxis function varies depending on the species of the microorganisms~\cite{12vincent1996}. For example, a typical phototaxis function is outlined as
\begin{equation}\label{5}
    M(G)=0.8\sin\left(\frac{3\pi}{2}\Lambda(G)\right)-0.1\sin\left(\frac{\pi}{2}\Lambda(G)\right),~~~where~~~ \Lambda(G)=Ge^{\beta(G-1)},
\end{equation}
    here $\beta$ is closely related to critical light intensity. 

\subsection{The governing equations}

    Assume that a continuous distribution is used to model cell population, as has been done in previous studies. In an incompressible dilute algal suspension, each algal cell has a volume $V$ and a density of $\rho+\Delta\rho$, where $\rho$ is the density of water. In this model, it is assumed that the all physical properties of the fluid are constant except the buoyancy force. The governing equations for the system are given by the following equations\\
    1. Continuity equation,
\begin{equation}\label{6}
	\boldsymbol{\nabla}\cdot \boldsymbol{u}=0,
\end{equation}
   where $\boldsymbol{u}$ is fluid velocity.\\
   2. In the rotating medium, the momentum equation under Boussinsque approximation
\begin{equation}\label{7}
	\rho\left(\frac{\partial \boldsymbol{u}}{\partial t}+(\boldsymbol{u}\cdot\nabla )\boldsymbol{u}+2\boldsymbol{\Omega}\times \boldsymbol{u}\right)=-\boldsymbol{\nabla} P_e+\mu\nabla^2\boldsymbol{u}-nVg\Delta\rho\hat{\boldsymbol{z}},
\end{equation}
    where $g$ is the gravitational acceleration due to gravity, $\boldsymbol{\Omega}=\Omega\hat{\boldsymbol{z}}$ is the angular velocity, $P_e$ is the excess pressure above hydrostatic, and $\mu$ is the viscosity of the fluid.\\
    3. Cell conservation equation
\begin{equation}\label{8}
	\frac{\partial n}{\partial t}=-\boldsymbol{\nabla}\cdot \boldsymbol{F_1}=\nabla\cdot[n\boldsymbol{u}+nW_c<\boldsymbol{p}>-\boldsymbol{D}\boldsymbol{\nabla} n].
\end{equation}
    Here, two key assumptions are made. First, the microorganisms are purely phototactic, and second, $\boldsymbol{ D}=DI$. With the help of these two assumptions, we can remove the Fokker-Plank equation from the governing system.~\cite{15panda2016}

    In this model, the lower horizontal boundary is assumed to be rigid and the upper horizontal boundary is assumed to be stress-free. Therefore, the boundary conditions can be expressed as
\begin{equation}\label{9}
	\boldsymbol{u}\cdot\hat{\boldsymbol{z}}=0\qquad at\quad z=0,H,
\end{equation}
\begin{equation}\label{10}
	\boldsymbol{F_1}\cdot\hat{\boldsymbol{z}}=0\qquad at\quad z=0,H,
\end{equation}
\begin{equation}\label{11}
	\boldsymbol{u}\times\hat{\boldsymbol{z}}=0\qquad at\quad z=0,
\end{equation}
\begin{equation}\label{12}
	\frac{\partial^2}{\partial z^2}(\boldsymbol{u}\cdot\hat{\boldsymbol{z}})=0\qquad at\quad z=H.
\end{equation}

    The top boundary is assumed to be exposed to collimated direct irradiation, then the boundary conditions for intensities are as follows
\begin{subequations}
	\begin{equation}\label{13a}
		at~~~ z=H,~~~~~~~	L(x, y, z , \theta, \phi)=L_t\delta(\boldsymbol{s}-\boldsymbol{s_0}),~~~ where~~~ (\pi/2\leq\theta\leq\pi),
	\end{equation}
	\begin{equation}\label{13b}
		at~~~ z=0,~~~~~~~ 	L(x, y, z , \theta, \phi) =0,~~~ where~~~ (0\leq\theta\leq\pi/2).
	\end{equation}
\end{subequations}	

\subsection{DIMENSIONLESS EQUATIONS}
     The equations that govern the system are transformed into a dimensionless form by selecting suitable scales for length ($H$), time ($H^2/D$), velocity ($D/H$), pressure ($\mu D/H^2$), and concentration ($\tilde{n}$). This is done to simplify the equations and make them easier to solve. The resulting dimensionless equations are presented below
\begin{equation}\label{14}
	\boldsymbol{\nabla}\cdot\boldsymbol{u}=0,
\end{equation}
\begin{equation}\label{15}
	S_c^{-1}\left(\frac{\partial \boldsymbol{u}}{\partial t}+(\boldsymbol{u}\cdot\nabla )\boldsymbol{u}\right)+\sqrt{T_a}(\hat{z}\times\boldsymbol{u})=-\nabla P_{e}-Rn\hat{\boldsymbol{z}}+\nabla^{2}\boldsymbol{u},
\end{equation}
\begin{equation}\label{16}
	\frac{\partial{n}}{\partial{t}}=-\boldsymbol{\nabla}\cdot\boldsymbol{F_1}=-{\boldsymbol{\nabla}}\cdot[\boldsymbol{n{\boldsymbol{u}}+nV_{c}<{\boldsymbol{p}}>-{\boldsymbol{\nabla}}n.}]
\end{equation}

    In the above equations, $S_c^{-1}=\nu/D$ represents the Schmidt number, $V_c$ denotes the dimensionless swimming speed as $V_c=W_cH/D$, $R=\tilde{n}V g\Delta{\rho}H^{3}/\nu\rho{D}$ is the Rayleigh number, and $T_a=4\Omega^2H^4/\nu^2$ is the Taylor number.

    After non-dimensionalization, the boundary conditions are expressed as
\begin{equation}\label{17}
	\boldsymbol{u}\cdot\hat{\boldsymbol{z}}=0\qquad at\quad z=0,1,
\end{equation}
\begin{equation}\label{18}
	\boldsymbol{F_1}\cdot\hat{\boldsymbol{z}}=0\qquad at\quad z=0,1,
\end{equation}
\begin{equation}\label{19}
	\boldsymbol{u}\times\hat{\boldsymbol{z}}=0\qquad at\quad z=0,
\end{equation}
\begin{equation}\label{20}
	\frac{\partial^2}{\partial z^2}(\boldsymbol{u}\cdot\hat{\boldsymbol{z}})=0\qquad at\quad z=1.
\end{equation}

    The RTE in dimensionless form is
\begin{equation}\label{21}
	\frac{dL}{dr}+\kappa nL(\boldsymbol{x},\boldsymbol{r})=0,
\end{equation}
    where $\kappa=\alpha\tilde{n}H$ is the dimensionless absorption coefficient. In dimensionless form, the boundary conditions for the intensity are
\begin{subequations}
\begin{equation}\label{22a}
	at~~~ z=1,~~~~~~~	L(x, y, z, \theta, \phi)=L_t\delta(\boldsymbol{r}-\boldsymbol{r_0}),~~~ where~~~ (\pi/2\leq\theta\leq\pi),
\end{equation}
\begin{equation}\label{22b}
	at~~~ z=0,~~~~~~~	L(x, y, z, \theta, \phi) =0,~~~ where~~~ (0\leq\theta\leq\pi/2). 
\end{equation}
\end{subequations}

\section{THE STEADY SOLUTION}

    The equations $(\ref{14})-(\ref{16})$ and $(\ref{21})$ have an equilibrium solution that can be described by the following equation

\begin{equation}\label{23}
	\boldsymbol{u}=0,~~~\zeta_s=\nabla\times u=0,~~~n=n_s(z)\quad and\quad  L=L_s(z,\theta),
\end{equation}
    where, $\zeta_s$, is the vorticity vector at the basic steady state.
    Thus, in the equilibrium state, the total intensity $G_s$ and radiative heat flux $\boldsymbol{q}_s$ can be expressed as follows
\begin{equation*}
	G_s=\int_0^{4\pi}L_s(z,\theta)d\Omega,\quad 
	\boldsymbol{q}_s=\int_0^{4\pi}L_s(z,\theta)\boldsymbol{s}d\Omega,
\end{equation*}
    and the governing equation for $L_s$ can be written as
\begin{equation}\label{24}
	\frac{dL_s}{dz}+\frac{\kappa n_sL_s}{\cos\theta}=0,
\end{equation}

    with the boundary conditions 
\begin{equation}\label{25}
	at ~~z=1,~~~~~L_s^c( 1, \theta) =L_t\delta(\boldsymbol{r}-\boldsymbol{r}_0),~~~ where~~~ (\pi/2\leq\theta\leq\pi), 
\end{equation}
    After calculations, we get 
\begin{equation}\label{26}
	L_s=L_t\exp\left(\int_z^1\frac{\kappa n_s(z')}{\cos\theta}dz'\right)\delta(\boldsymbol{r}-\boldsymbol{r_0}), 
\end{equation}

    Now the total intensity, $G_s$ in the equilibrium state, can be written as
\begin{equation}\label{27}
	G_s=\int_0^{4\pi}L_s(z,\theta)d\Omega=L_t\exp\left(-\int_z^1\frac{\kappa n_s(z')dz'}{\cos\theta_0}\right),
\end{equation}

    The radiative flux in the basic state can be expressed as follows

\begin{equation*}
	\boldsymbol{q_s}=-L_t\exp\left(-\int_z^1\frac{\kappa n_s(z')dz'}{\cos\theta_0}\right)\cos\theta_0\hat{\boldsymbol{z}}.
\end{equation*}
    The mean swimming direction can be determined as follows,
\begin{equation*}
	<\boldsymbol{p_s}>=M_s\hat{\boldsymbol{z}},
\end{equation*}

    where $M_s=M(G_s).$\par
    The solution for the cell concentration in a basic state can be expressed as $n_s(z)$ and satisfies the following equation

\begin{equation}\label{28}
	\frac{dn_s}{dz}-V_cM_sn_s=0,
\end{equation}
    where, the basic state cell concentration $n_s(z)$ is accompanied by the conservation relation for the cells
\begin{equation}\label{29}
	\int_0^1n_s(z)dz=1.
\end{equation}
    The equations given by (\ref{28}) and (\ref{29}) forms a boundary value problem, and the solution to this problem is obtained using a numerical technique called the shooting method.

\section{Linear stability of the problem}
    For stability analysis, we use linear perturbation theory. Here, the small perturbation of amplitude ${\bar{\epsilon}} (0<{\bar{\epsilon}}<<1)$ is made to the equilibrium state, according to the following equation

	\begin{align}\label{37}
	\nonumber[\boldsymbol{u},\zeta,n,L,<p>]=[0,\zeta_s,n_s,L_s,<p_s>]+\bar{\epsilon} [\boldsymbol{u}_1,\zeta_1,n_1,L_1,<\boldsymbol{p}_1>]+\mathcal{O}(\bar{\epsilon}^2).  
	\end{align}

    Eqs.~(\ref{14})-(\ref{16}) are linearized by substituting the perturbed variables and collecting $o(\bar{\epsilon})$ terms about the equilibrium state, gives
\begin{equation}\label{30}
	\boldsymbol{\nabla}\cdot \boldsymbol{u}_1=0,
\end{equation}
    where  $\boldsymbol{u}_1=(u_1,v_1,w_1)$.
\begin{equation}\label{31}
	S_c^{-1}\left(\frac{\partial \boldsymbol{u_1}}{\partial t}\right)+\sqrt{T_a}(z\times u_1)+\boldsymbol{\nabla} P_{e}+Rn_1\hat{\boldsymbol{z}}=\nabla^{2}\boldsymbol{ u_1},
\end{equation}
\begin{equation}\label{32}
	\frac{\partial{n_1}}{\partial{t}}+V_c\boldsymbol{\nabla}\cdot(<\boldsymbol{p_s}>n_1+<\boldsymbol{p_1}>n_s)+w_1\frac{dn_s}{dz}=\boldsymbol{\nabla}^2n_1.
\end{equation}
   If $G=G_s+\bar{\epsilon}G_1+\mathcal{O}(\bar{\epsilon}^2)$, then the total intensity in the basic state is perturbed as $L_t\exp\left(\frac{-\kappa\int_z^1(n_s(z')+\bar{\epsilon} n_1+\mathcal{O}(\bar{\epsilon}^2))dz'}{\cos\theta_0}\right)$  and after simplification, we get
\begin{equation}\label{33}
	G_1=L_t\exp\left(\frac{\int_1^z \kappa n_s(z')dz'}{\cos\theta_0}\right)\left(\frac{\int_1^z\kappa n_1 dz'}{\cos\theta_0}\right),
\end{equation}
    Hence, the perturbed mean swimming orientation [i.e., $T(G)\hat{\boldsymbol{z}}-T_s(G)\hat{\boldsymbol{z}}$] at $\mathcal{O}(\bar{\epsilon}^2)$ for a non-scattering algal suspension is expressed as
\begin{equation}\label{34}
	<p_1>=G_1\frac{dT_s}{dG}\hat{\boldsymbol{z}}
\end{equation}

    We eliminate $P_e$ and the horizontal component of $u_1$ by taking
    the curl of Eq.~(\ref{31}) twice and retaining the z-component of the result. Then, Eqs.~(\ref{30})–(\ref{32}) reduce to three equations for the perturbed variables, namely the vertical component of the velocity $w_1$, the vertical component of the vorticity $\zeta_1 (= \zeta\cdot\hat{\boldsymbol{z}})$ and the concentration $n_1$. These variables can be decomposed into normal modes as
\begin{equation}\label{35}
	[w_1,\zeta_1,n_1]=[W(z),Z(z),N(z)]\exp{(\sigma t+i(lx+my))}.  
\end{equation}

	The linear stability equations become
\begin{equation}\label{36}
	\left(\sigma S_c^{-1}+k^2-D^2\right)\left( D^2-k^2\right)W(z)=Rk^2N,
\end{equation}
\begin{equation}\label{37}
	\left(\sigma S_c^{-1}+k^2-D^2\right)Z(z)=\sqrt{T_a}DW(z)
\end{equation}
\begin{equation}\label{38}
	\aleph_1(z)\int_z^1N(z') dz'+(\sigma+k^2+\aleph_2(z))N(z)+\aleph_3(z)DN(z)-D^2N(z)=-Dn_sW(z), 
\end{equation}
	where
\begin{subequations}
		
\begin{equation}\label{39a}
	\aleph_1(z)=-(\kappa/\cos\theta_0) V_cD\left(n_sG_s\frac{dM_s}{dG}\right),
\end{equation}
\begin{equation}\label{39b}
	\aleph_2(z)=2(\kappa/\cos\theta_0) V_c n_s G_s\frac{dM_s}{dG},
\end{equation}
\begin{equation}\label{39c}
	\aleph_3(z)=V_cM_s.
\end{equation}
\end{subequations}
	The boundary conditions become,
\begin{equation}\label{40}
	at~~z=0,~~W(z)=\frac{dW(z)}{dz}=Z(z)=\frac{dN(z)}{dz}-\aleph_3(z)N(z)+n_sV_c(\kappa/\cos\theta_0)G_s\left(\int_z^1N(z')dz'\right)\frac{dM_s}{dG}=0.
\end{equation}
\begin{equation}\label{41}
	at~~z=1,~~W(z)=\frac{dW(z)}{dz}=\frac{dZ(z)}{dz}=\frac{dN(z)}{dz}-\aleph_3(z)N(z)=0.
\end{equation}
	
	Here, $k=\sqrt{(l^2+m^2)}$ is the non-dimensional wavenumber.
	
	Now, introducing a new variable as
\begin{equation}\label{42}
	\Phi(z)=\int_1^zN(z')dz',
\end{equation}
	Eqs.~\ref{36}-\ref{38} becomes
\begin{equation}\label{43}
	\left(\sigma S_c^{-1}+k^2-D^2\right)\left( D^2-k^2\right)W=Rk^2D\Phi,
\end{equation}
	\begin{equation}\label{44}
\left(\sigma S_c^{-1}+k^2-D^2\right)Z(z)=\sqrt{T_a}DW
\end{equation}
\begin{equation}\label{45}
	\aleph_1(z)\Phi+(\sigma+k^2+\aleph_2(z))D\Phi+\aleph_3(z)D^2\Phi-D^3\Phi=-Dn_sW, 
\end{equation}
	with the boundary conditions,
	
\begin{equation}\label{46}
	at~~z=0,~~W=DW=Z(z)=D^2\Phi-\aleph_3(z)\Phi-n_sV_c(\kappa/\cos\theta_0)G_s\Phi\frac{dM_s}{dG}=0.
\end{equation}
\begin{equation}\label{47}
	at~~z=1,~~W=DW=DZ(z)=D^2\Phi-\aleph_3(z)\Phi=0,
\end{equation}

	and the additional boundary condition is,
\begin{equation}\label{48}
	at~~~z=1,~~~~~~\Phi(z)=0.
\end{equation}

\section{SOLUTION technique}
To find solutions for Eqs. (\ref{43}) to (\ref{45}), we employ the Newton-Raphson-Kantorovich (NRK) iterative method, as described by Cash et al.~\cite{19cash1980}. This numerical approach allows us to calculate the growth rate, Re$(\sigma)$, or neutral stability curves in the $(k, R)$-plane for a specific set of parameters. The neutral curve, denoted as $R^{(n)}(k)$, where $n$ is an integer greater than or equal to 1, consists of an infinite number of branches. Each branch represents a possible solution to the linear stability problem for the given parameter set. Among these branches, the one with the lowest value of $R$ is considered the most significant, and the corresponding bioconvective solution is identified as $(k_c, R_c)$. This particular solution is referred to as the most unstable solution. By utilizing the equation $\lambda_c=2\pi/k_c$, where $\lambda_c$ represents the wavelength of the initial disturbance, we can determine the wavelength associated with the most unstable solution. This wavelength provides valuable information about the characteristic pattern size of the bioconvection phenomenon.
	
\section{NUMERICAL RESULTS}
	
    In our study, we recognize the complexity of exploring the entire parameter space due to the wide range of values that each parameter can take. To systematically investigate the impact of rotation, represented by the Taylor number $T_a$, we decided to keep certain parameters constant while varying others. This approach allows us to focus on specific aspects of the system and gain a deeper understanding of their individual influence on the onset of bioconvection.
    
    Throughout the study, we fix $S_c=20$ and $L_t=0.8$ as constant parameter values. These choices are made to maintain consistency and isolate the effects of other parameters. The parameters related to the absorption coefficient and cell swimming speed, namely $\kappa$ and $V_c$, are varied to observe their specific impacts. We consider two values for $\kappa$, namely 0.5 and 1.0, representing different light absorption characteristics of the microorganisms. Additionally, we explore three values for $V_c$, specifically 10, 15, and 20, to study how the swimming speed of the cells influences the bioconvection behavior. Furthermore, we investigate the angle of incidence, $\theta_i$, over a range from 0° to 80°. By considering different angles of incidence, we can observe how the orientation of light influences the initiation of bioconvection.

\subsection{$V_c=$10}
	
   In this section, our focus centers on investigating the combined effect of two crucial factors, the Taylor number denoted as $T_a$, and the angle of incidence $\theta_i$, on the intriguing bioconvective instability. We carefully choose three specific angles of incidence, $\theta_i=0°, 40°,$ and $80°$, to explore their impact on the system. Additionally, the critical intensity $G_c$ is deliberately selected to position the sublayer at the mid-height of the suspension for $\theta_i=0°$. As we progressively increase the angle of incidence to 40° and 80°, the sublayer's location shifts to approximately three-quarters height and finally to the top of the suspension.
   To comprehensively observe the resulting changes in the bioconvective instability, we systematically vary the value of $T_a$ across a wide range, from lower to higher values, reaching up to 10,000. By exploring different values of $T_a$, we aim to understand how the presence of rotation influences the onset and characteristics of bioconvection at different locations within the suspension. In our investigation, we consider two distinct cases, differing in the value of $\kappa$, the extinction coefficient. This choice allows us to analyze how the variation in the extinction coefficient influences the bioconvective behavior under the influence of rotation.
    
\subsubsection{$\kappa=0.5$}

\begin{figure*}[!ht]
	\includegraphics{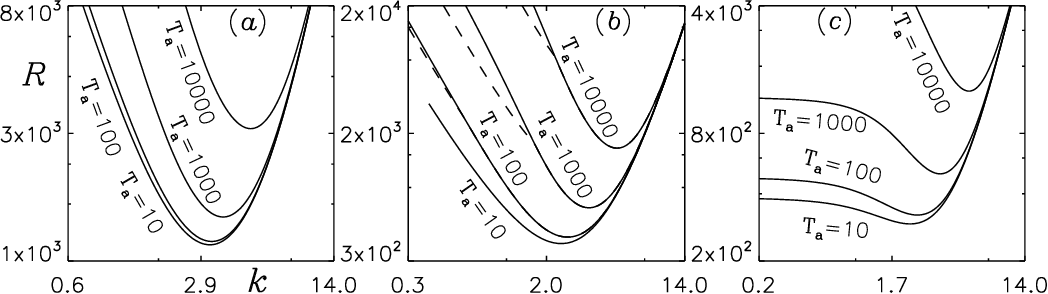}
	\caption{\label{fig2} The marginal stability curves (a) for $\theta_i=0$, (b) $\theta_i=40$, and (c) $\theta_i=80$. Here, the parameter values $S_c=20,V_c=10, \kappa=0.5$, and $G_c=0.63$ are kept fixed.}
\end{figure*}

In Fig.~\ref{fig2}(a), we present the marginal stability curves for $\theta_i=0°$, with constant values $V_c=10$, $\kappa=0.5$, and $G_c=0.63$, while varying the Taylor number ($T_a$) at different levels.
When $T_a=10$, the linear stability analysis predicts a stationary solution at the bioconvective instability with a finite pattern wavelength. As we increase the Taylor number to 100, the solution remains stationary, and the critical Rayleigh number increases, while the pattern wavelength decreases. Continuing to $T_a=1000$ and up to 10000, the marginal stability curves exhibit a similar behavior, where the critical Rayleigh number continues to increase, and the pattern wavelength decreases.

Fig.~\ref{fig2}(b) shows the marginal stability curve for $\theta_i=40°$. When $T_a=10$, the linear stability analysis predicts a stationary solution at the bioconvective instability with a finite pattern wavelength. However, as we increase the Taylor number to 100, an oscillatory branch bifurcates from the stationary branch of the marginal stability curve. Despite this, the most unstable solution still occurs at the stationary branch, leading to a stationary solution. The critical Rayleigh number increases, and the pattern wavelength decreases. Similar behavior is observed for $T_a=1000$ and up to 10000, where the marginal stability curves continue to exhibit an oscillatory branch, but the most unstable solution remains on the stationary branch, resulting in a stationary solution, while the critical Rayleigh number keeps increasing and the pattern wavelength decreases.

In Fig.~\ref{fig2}(c), the marginal stability curve for $\theta_i=80°$ is shown. Similar to the previous cases, when $T_a=10$, the linear stability analysis predicts a stationary solution at the bioconvective instability with a finite pattern wavelength. As we increase the Taylor number to 100, the solution remains stationary, and the critical Rayleigh number increases, while the pattern wavelength decreases. Continuing to $T_a=1000$ and up to 10000, the marginal stability curves exhibit a similar behavior, where the critical wavelength decreases, and the critical Rayleigh number increases.

For a comprehensive overview of our numerical findings, Table~\ref{tab1} presents the results obtained in this section.

\begin{table}[!htbp]
	\caption{\label{tab1}The table shows the numerical results of bioconvective solutions for different values of Taylor number $T_a$, where the other governing values $V_c=$10, $\kappa=$0.5, $G_c=0.63$, and $\theta_i=0,40,80$ are keep constant. A result with double dagger symbol indicates that a smaller minimum occurs on an oscillatory branch, and a starred result indicates that $R^{(1)}(k)$ a branch of the neutral curve is oscillatory.}
\begin{ruledtabular}
	\begin{tabular}{cccccccccc}
	\multirow{2}{*}{$T_a$}&\multicolumn{3}{c}{$\theta_i=0$} & \multicolumn{3}{c}{$\theta_i=40$} & \multicolumn{3}{c}{$\theta_i=80$}\\\cline{2-4}\cline{5-7}\cline{8-10} & $\lambda_c$ & $R_c$ & $Im(\sigma)$ &$ \lambda_c$ & $R_c$ & $Im(\sigma)$ &$\lambda_c$ & $R_c$ & $Im(\sigma)$\\
	\hline
	10 & 1.94 & 1353.28 & 0 & 2.51 & 396.57 & 0 & 2.83 & 255.30 & 0 \\ 	
	100 & 1.91 & 1387.21 & 0 & 2.27$^{\star}$ & 439.94 & 0 & 2.51 & 285.54 & 0 \\ 
	1000 & 1.68 & 1664.03 & 0 & 1.69$^{\star}$ & 704.52 & 0 & 1.73 & 480.29 & 0 \\ 
	10000 & 1.21 & 3209.80 & 0 & 1.14$^{\star}$ & 1833.71 & 0 & 1.10 & 1363.77 & 0 
		
	\end{tabular}
\end{ruledtabular}
	
\end{table}

%%%%%%%%%%%%%%%%%%%%%%%%%%%%%%%%%%%%%%%%%%%%%%%%%%%%%%%%%%%%%%%%%%%%%%%%%%

\subsubsection{$\kappa=1$}

\begin{figure*}[!ht]
	\includegraphics{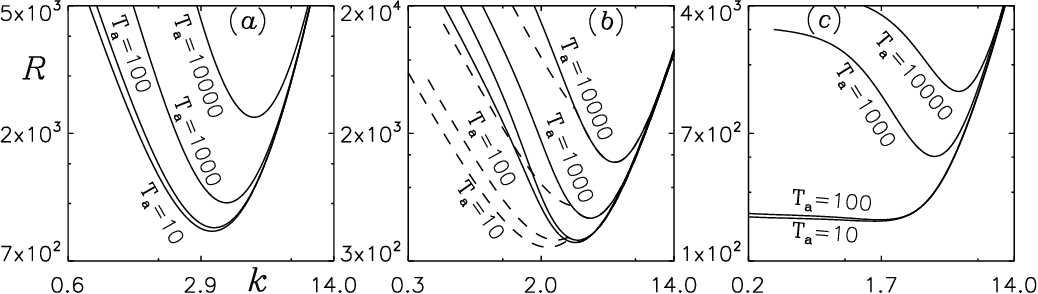}
	\caption{\label{fig3}The marginal stability curves (a) for $\theta_i=0$, (b) $\theta_i=40$, and (c) $\theta_i=80$. Here, the parameter values $S_c=20,V_c=10, \kappa=1$, and $G_c=0.495$ are kept fixed.}
\end{figure*}

In this section, we present the results obtained for $V_c=10$, $\kappa=1$, and $G_c=0.495$. Fig.~\ref{fig3}(a) shows the marginal stability curves for $\theta_i=0°$, where we vary the Taylor number ($T_a$) while keeping the other parameters constant.

When $T_a=10$, the linear stability analysis predicts a stationary solution at the bioconvective instability with a finite pattern wavelength. As we increase the Taylor number to 100, the solution remains stationary, and the critical Rayleigh number increases, while the pattern wavelength decreases. Continuing to $T_a=1000$ and up to 10000, the marginal stability curves exhibit a similar behavior, where the critical Rayleigh number continues to increase, and the pattern wavelength decreases.

Fig.~\ref{fig3}(b) presents the marginal stability curve for $\theta_i=40°$. When $T_a=10$, an oscillatory branch bifurcates from the stationary branch of the marginal stability curve, and the most unstable solution occurs at the oscillatory branch, leading to an oscillatory solution with a finite pattern wavelength. As we increase the Taylor number to 100, an oscillatory branch bifurcates again from the stationary branch of the marginal stability curve, and the most unstable solution occurs at the oscillatory branch, resulting in an oscillatory solution. Here, the critical Rayleigh number increases, but the pattern wavelength decreases as $T_a$ increases. However, as we set $T_a=1000$, an interesting change occurs, where an oscillatory branch bifurcates once more, but the most unstable solution now occurs at the stationary branch of the marginal stability curve, leading to a stationary solution. This means that, for $T_a=1000$, the nature of the bioconvective instability changes from oscillatory to stationary. A similar nature of bioconvective instability is observed for $T_a=10000$, where the critical Rayleigh number increases while the critical pattern wavelength decreases.

Fig.~\ref{fig3}(c) displays the marginal stability curve for $\theta_i=80°$. Similar to the previous cases, when $T_a=10$, the linear stability analysis predicts a stationary solution at the bioconvective instability with a finite pattern wavelength. As we increase the Taylor number to 100, the solution remains stationary, and the critical Rayleigh number increases, while the pattern wavelength decreases. Continuing to $T_a=1000$ and up to 10000, the marginal stability curves exhibit a similar behavior, where the critical wavelength decreases, and the critical Rayleigh number increases.

For a comprehensive summary of our numerical findings, Table~\ref{tab2} provides the results obtained in this section.

\begin{table}[!htbp]
	\caption{\label{tab2}The table shows the numerical results of bioconvective solutions for different values of Taylor number $T_a$, where the other governing values $V_c=$10, $\kappa=$1, $G_c=0.495$, and $\theta_i=0,40,80$ are keep contant. A result with double dagger symbol indicates that a smaller minimum occurs on an oscillatory branch, and a starred result indicates that $R^{(1)}(k)$ a branch of the neutral curve is oscillatory.}
\begin{ruledtabular}
	\begin{tabular}{cccccccccc}
		\multirow{2}{*}{$T_a$}&\multicolumn{3}{c}{$\theta_i=0$} & \multicolumn{3}{c}{$\theta_i=40$} & \multicolumn{3}{c}{$\theta_i=80$}\\\cline{2-4}\cline{5-7}\cline{8-10} & $\lambda_c$ & $R_c$ & $Im(\sigma)$ &$ \lambda_c$ & $R_c$ & $Im(\sigma)$ &$\lambda_c$ & $R_c$ & $Im(\sigma)$\\
		\hline
		10 & 1.89 & 880.17 & 0 & 2.78$^{\ddag}$ & 373.94$^{\ddag}$ & 6.26 & 4.23 & 221.77 & 0 \\ 		
		100 & 1.86 & 904.38 & 0 & 2.59$^{\ddag}$ & 419.31$^{\ddag}$ & 6.15 & 3.06 & 257.05 & 0 \\
		1000 & 1.59 & 1095.77 & 0 & 1.50$^{\star}$ & 610.21 & 0 & 1.81 & 451.72 & 0 \\ 
		10000 & 1.16 & 2117.94 & 0 & 1.07$^{\star}$ & 1484.77 & 0 & 1.09 & 1266.85 & 0 
	\end{tabular}
\end{ruledtabular}
	
\end{table}

%%%%%%%%%%%%%%%%%%%%%%%%%%%%%%%%%%%%%%%%%%%%%%%%%%%%%%%%%%%%%%%%%%%%%%%%%%	

\subsection{$V_c=$15 and $V_c=$20}
We have also investigated the effect of rotation (i.e. Taylor number) and oblique irradiation on bio-convective instability for $V_c=15$ and $V_c=20$. The critical Rayleigh number, $R_c$, and wavelength, $\lambda_c$, obtained from the numerical simulations are shown in Tables~\ref{tab3},\ref{tab4},\ref{tab5},\ref{tab6}.   

\begin{table}[!htbp]
	\caption{\label{tab3}The table shows the numerical results of bioconvective solutions for different values of Taylor number $T_a$, where the other governing values $V_c=$15, $\kappa=$0.5, $G_c=0.63$, and $\theta_i=0,40,80$ are keep constant. A result with double dagger symbol indicates that a smaller minimum occurs on an oscillatory branch, and a starred result indicates that $R^{(1)}(k)$ a branch of the neutral curve is oscillatory.}
	\begin{ruledtabular}
		\begin{tabular}{cccccccccc}
			\multirow{2}{*}{$T_a$}&\multicolumn{3}{c}{$\theta_i=0$} & \multicolumn{3}{c}{$\theta_i=40$} & \multicolumn{3}{c}{$\theta_i=80$}\\\cline{2-4}\cline{5-7}\cline{8-10} & $\lambda_c$ & $R_c$ & $Im(\sigma)$ &$ \lambda_c$ & $R_c$ & $Im(\sigma)$ &$\lambda_c$ & $R_c$ & $Im(\sigma)$\\
			\hline
			10 & 1.89 & 827.27 & 0 & 2.71$^{\ddag}$ & 332.34$^{\ddag}$ & 9.44 & 1.78$^{\star}$ & 450.75 & 0 \\ 		
			100 & 1.86 & 850.55 & 0 & 2.53$^{\ddag}$ & 371.27$^{\ddag}$ & 9.63 & 1.72$^{\star}$ & 469.45 & 0 \\ 
			1000 & 1.59 & 1032.56 & 0 & 1.41$^{\star}$ & 639.43 & 0 & 1.45 & 610.28 & 0 \\ 
			10000 & 1.15 & 1997.97 & 0 & 1.03$^{\star}$ & 1454.33 & 0 & 1.01 & 1302.74 & 0 \\ 
			
		\end{tabular}
	\end{ruledtabular}
	
\end{table}

\begin{table}[!htbp]
	\caption{\label{tab4}The table shows the numerical results of bioconvective solutions for different values of Taylor number $T_a$, where the other governing values $V_c=$15, $\kappa=$1, $G_c=0.5$, and $\theta_i=0,40,80$ are keep constant. A result with double dagger symbol indicates that a smaller minimum occurs on an oscillatory branch, and a starred result indicates that $R^{(1)}(k)$ a branch of the neutral curve is oscillatory.}
	\begin{ruledtabular}
		\begin{tabular}{cccccccccc} 
			\multirow{2}{*}{$T_a$}&\multicolumn{3}{c}{$\theta_i=0$} & \multicolumn{3}{c}{$\theta_i=40$} & \multicolumn{3}{c}{$\theta_i=80$}\\\cline{2-4}\cline{5-7}\cline{8-10} & $\lambda_c$ & $R_c$ & $Im(\sigma)$ &$ \lambda_c$ & $R_c$ & $Im(\sigma)$ &$\lambda_c$ & $R_c$ & $Im(\sigma)$\\
			\hline
			10 & 1.77 & 446.15 & 0 & 2.42$^{\ddag}$ & 335.83$^{\ddag}$ & 18.30 & 4.91 & 279.87 & 0 \\ 		
			100 & 1.74 & 663.93 & 0 & 2.27$^{\ddag}$ & 368.52$^{\ddag}$ & 18.69 & 3.21 & 321.60 & 0 \\ 
			1000 & 1.48 & 800.91 & 0 & 1.77$^{\ddag}$ & 604.50$^{\ddag}$ & 19.76 & 1.78 & 520.39 & 0 \\ 
			10000 & 1.08 & 1497.86 & 0 & 0.93$^{*}$ & 1488.21 & 0 & 1.05 & 1231.81 & 0
			
		\end{tabular}
	\end{ruledtabular}
	
\end{table}

%%%%%%%%%%%%%%%%%%%%%%%%%%%%%%%%%%%%%%%%%%%%%%%%%%%%%%%%%%%%%%%%%%%%%%%%
	
\begin{table}[!htbp]
	\caption{\label{tab5}The table shows the numerical results of bioconvective solutions for different values of Taylor number $T_a$, where the other governing values $V_c=$20, $\kappa=$0.5, $G_c=0.63$, and $\theta_i=0,40,80$ are keep constant. A result with double dagger symbol indicates that a smaller minimum occurs on an oscillatory branch, and a starred result indicates that $R^{(1)}(k)$ a branch of the neutral curve is oscillatory.}
	\begin{ruledtabular}
		\begin{tabular}{cccccccccc} 
			\multirow{2}{*}{$T_a$}&\multicolumn{3}{c}{$\theta_i=0$} & \multicolumn{3}{c}{$\theta_i=40$} & \multicolumn{3}{c}{$\theta_i=80$}\\\cline{2-4}\cline{5-7}\cline{8-10} & $\lambda_c$ & $R_c$ & $Im(\sigma)$ &$ \lambda_c$ & $R_c$ & $Im(\sigma)$ &$\lambda_c$ & $R_c$ & $Im(\sigma)$\\
			\hline
			10 & 1.80 & 676.31 & 0 & 2.42$^{\ddag}$ & 324.61$^{\ddag}$ & 18.30 & 1.90$^{\ddag}$ & 279.87$^{\ddag}$ & 12.86 \\ 		
			100 & 1.74 & 694.58 & 0 & 2.32$^{\ddag}$ & 356.66$^{\ddag}$ & 19.10 & 1.33$^{*}$ & 321.60 & 0 \\ 
			1000 & 1.50 & 836.29 & 0 & 1.80$^{\ddag}$ & 588.06$^{\ddag}$ & 20.56 & 1.22$^{*}$ b & 520.39 & 0 \\ 
			10000 & 1.09 & 1567.89 & 0 & 0.93$^{*}$ & 1513.03 & 0 & 0.94 & 1231.81 & 0
			
		\end{tabular}
	\end{ruledtabular}
	
\end{table}

\begin{table}[!htbp]
	\caption{\label{tab6}The table shows the numerical results of bioconvective solutions for different values of Taylor number $T_a$, where the other governing values $V_c=$20, $\kappa=$1, $G_c=0.49$, and $\theta_i=0,40,80$ are keep constant.A result with double dagger symbol indicates that a smaller minimum occurs on an oscillatory branch, and a starred result indicates that $R^{(1)}(k)$ a branch of the neutral curve is oscillatory.}
	\begin{ruledtabular}
		\begin{tabular}{cccccccccc} 
			\multirow{2}{*}{$T_a$}&\multicolumn{3}{c}{$\theta_i=0$} & \multicolumn{3}{c}{$\theta_i=40$} & \multicolumn{3}{c}{$\theta_i=80$}\\\cline{2-4}\cline{5-7}\cline{8-10} & $\lambda_c$ & $R_c$ & $Im(\sigma)$ &$ \lambda_c$ & $R_c$ & $Im(\sigma)$ &$\lambda_c$ & $R_c$ & $Im(\sigma)$\\
			\hline
			10 & 1.69 & 495.21 & 0 & 2.18$^{\ddag}$ & 413.55$^{\ddag}$ & 31.84 & 4.91 & 352.64 & 0 \\ 		
			100 & 1.64 & 611.12 & 0 & 2.06$^{\ddag}$ & 494.60$^{\ddag}$ & 32.51 & 3.52 & 403.12 & 0 \\ 
			1000 & 1.41 & 731.10 & 0 & 1.61$^{\ddag}$ & 658.57$^{\ddag}$ & 35.06 & 1.80 & 632.61 & 0 \\ 
			10000 & 1.03 & 1312.86 & 0 & 1.08$^{*}$ & 1671.36 & 35.25 & 1.02 & 1332.21 & 0
			
		\end{tabular}
	\end{ruledtabular}
	
\end{table}	

%%%%%%%%%%%%%%%%%%%%%%%%%%%%%%%%%%%%%%%%%%%%%%%%%%%%%%%%%%%%%%%%%%%%%%%%%	
	
\section{Conclusion}
   In this study, we have made significant strides in understanding the complex behavior of bioconvection patterns by incorporating the combined influence of rotation and oblique collimated flux in a non-scattering suspension of phototactic microorganisms. This novel model, the first of its kind, allows us to explore the intricate dynamics that arise when microorganisms respond to varying light conditions and experience rotational effects.
   
   Our analysis has led to several important conclusions:
   \begin{enumerate}
   	\item The total intensity of light decreases as the angle of incidence increases from zero to non-zero values, due to the self-shading effect. This highlights the importance of considering oblique collimated flux to accurately represent natural lighting conditions.
   	
   	\item The location of the sublayer, representing the equilibrium state, shifts from the mid-height to the top of the suspension as the angle of incidence increases. This shift in the sublayer's position illustrates how microorganisms respond differently to light when exposed to varying angles of incidence.
   	
   	\item The linear stability analysis reveals the existence of both stationary (non-oscillatory) and oscillatory (non-stationary) solutions at the onset of bioconvective instability. The transition from oscillatory to stationary solutions occurs at specific Taylor numbers, indicating the critical role of rotation in determining the nature of bioconvection patterns.
   	
   	\item  The critical Rayleigh number increases, and the critical initial pattern wavelength decreases as the Taylor number (rotation rate) increases at every angle of incidence. This demonstrates how rotation affects the stability and pattern formation in the suspension.
   \end{enumerate}

   However, it is essential to acknowledge that validation of our model requires experimental data on purely phototactic bioconvection, which is currently limited due to the scarcity of suitable microorganisms. Future studies should focus on identifying microorganisms exhibiting pure phototactic behavior to validate and further refine our model. Moreover, our proposed model lays the foundation for simulating other phototactic bioconvection phenomena of interest, expanding our understanding of the interactions between microorganisms and light. The exploration of other light conditions, scattering effects, and more complex scenarios will pave the way for deeper insights into bioconvection dynamics and their implications in various biological and ecological contexts.

\section*{ Availability of Data}
	The supporting data of this article is available within the article. 
\nocite{*}
	%	\section*{REFERENCES}
\bibliography{OBLIQUE_ROTATION}
	
\end{document}